# Modeling Transport of SARS-CoV-2 Inside a Charlotte Area Transit System (CATS) Bus

**Gregory McGowan[1], Jeffrey Feaster[1], Andy Jones[1], Lucas Agricola[1], Matthew Goodson[1], William Timms[2], and Mesbah Uddin[2,*]**

[1]Corvid Technologies, Mooresville, NC 28117, USA
[2]UNC Charlotte, Charlotte, NC 28223, USA
[*]Corresponding Author: Mesbah.Uddin@uncc.edu

**ABSTRACT**

We present in this paper a model of the transport of human respiratory particles on a Charlotte Area Transit System (CATS) bus to examine the efficacy of interventions to limit exposure to SARS-CoV-2, the virus that causes COVID-19. The methods discussed here utilize a commercial Navier-Stokes flow solver, RavenCFD, run using a massively parallel supercomputer to model the flow of air through the bus under varying conditions, such as windows being open or the HVAC flow settings. Lagrangian particles are injected into the RavenCFD predicted flow fields to simulate the respiratory droplets from speaking, coughing, or sneezing. These particles are then traced over time and space until they interact with a surface or are removed via the HVAC system. Finally, a volumetric Viral Mean Exposure Time (VMET) is computed to quantify the risk of exposure to the SARS-CoV-2 under various environmental and occupancy scenarios. Comparing the VMET under varying conditions should help identify viable methods to reduce the risk of viral exposure of CATS bus passengers during the COVID-19 pandemic.

**Nomenclature**

| | | |
|---|---|---|
| $d_p$ | = | particle diameter |
| $P_f$ | = | fluid pressure |
| $T_f$ | = | fluid temperature |
| $\vec{u}_f$ | = | fluid velocity vector |
| $\vec{u}_p$ | = | particle velocity vector |
| $\vec{x}_p$ | = | particle location vector |
| $\alpha$ | = | drag coefficient |
| $\lambda$ | = | mean free path |
| $\mu_f$ | = | fluid dynamic viscosity |
| $\rho_f$ | = | fluid density |
| $\rho_p$ | = | particle density |
| $\tau_p$ | = | particle characteristic stopping time |
| $\text{Re}_p$ | = | particle Reynolds number |
| $\text{Kn}$ | = | Knudsen number |
| VMET | = | Viral Mean Exposure Time |

## 1 Introduction

Understanding and monitoring airflow is critical to mitigating the presence and spread of viral particles - especially in enclosed spaces such as office buildings, restaurants, gymnasiums, and public transportation systems. This effort aims to apply computational fluid dynamics (CFD) to characterize the airflow interior of a public transportation system and simulate modes of transmission such as breathing, sneezing, and coughing via Lagrangian particle tracking (LPT) methods which may be used to inform operational decisions. This research presents a joint effort by the University of North Carolina at Charlotte and Corvid Technologies to create a monitoring model of a Charlotte Area Transportation System (CATS) bus. This project is supported by the Coronavirus Aid, Relief, and Economic and Security Act (CARES Act) as part of an award from the North Carolina Pandemic Recovery Office. The contents are those of the author(s) and do not necessarily represent the official views of, nor an endorsement, by the State of North Carolina or the U.S. Government.

Given the internal structure of the vehicle, its ventilation systems, and the location of passengers, we aim to determine the likely distribution of viruses in the bus resulting from three common respiratory events: speaking, coughing, and sneezing. This allows for identification of any specific surfaces or locations where virus particles are most likely accumulate (given the flow dynamics). These areas can then be targeted for cleaning or passenger avoidance. The overarching goals of this study include:

- Develop first-principle based, high-fidelity, interior airflow datasets

- Develop an automated Lagrangian particle tracking (LPT) technique relevant to respiratory particle sizing

- Develop viral load models for various vehicle configurations and transmission mechanisms

Utilizing these datasets this research effort aims to address the following questions by developing a COVID-19 Public Transportation Monitoring Model that uses the data generated by hundreds of first-principles, physics-based Navier-Stokes Calculations:

- Are deposition location and particle density dependent on the distribution and density of passengers?

- Given the ventilation structure and the behavior of aerosol and droplet particles, where are the optimal locations for placement of air samplers in the vehicle to maximize the chance of detecting the virus?

- Assuming that new findings pointing to the greater-than-anticipated aerosol spread of SARS-CoV-2 are correct, to what extent do aerosol particles accumulate in the interior air while the vehicle is operating?

Throughout the COVID-19 pandemic the scientific community has been learning at an incredibly fast pace and has been met with significant informational challenges. Lack of information and inaccurate information (both intentional and accidental) can all create significant uncertainty. In order to shed light on viral transmission, this research effort relies on a first principles physics-based approach rather than a collection of controlled and uncontrolled field studies wrapped into a statistical representation. The number of variables associated with the problem of viral transmission are significant and case-specific which requires a research team to dedicate significant resources and rely heavily on automation in order to develop required number of datasets. Our approach is to model bulk flow patterns using Navier-Stokes methods, evaluate viral dispersion via Lagrangian particle tracking methods modified to include aerosols, and produce a viral load map which can be leveraged to identify probable areas of contamination throughout the control volume. Trends in the aggregate of solutions will provide insight into common areas of deposit which would be primary target areas for sanitation measures and/or avoidance, and possibly operational procedures and configuration changes (seating arrangement/HVAC settings).

This paper is structured as following:. A description of the problem setup and methods, including the bus model, Navier-Stokes flow solver, LPT algorithm, and viral load, are described in Sec. 2. Sec. 3 details the validation studies of the flow solver and the LPT algorithms. Sec. 4 presents preliminary results of the study, which are finally summarized in Sec. 5.

## 2 Setup and Methods

### 2.1 Flow Solver

Simulations are performed using RavenCFD, a proprietary Navier-Stokes solver by Corvid Technologies. RavenCFD solves the Reynolds-Averaged Navier-Stokes (RANS) equations on an unstructured, arbitrary polyhedral mesh. Simulations are performed using fully implicit time integration based on the methods of Ref. 1, 2. Gradients are computed using a weighted least-squares approach and limited using the minmod algorithm (e.g., Ref. 3). Face values are reconstructed using second order interpolation from cell centers. Inviscid fluxes are computed using the all-speed Low-Diffusion Flux Splitting Scheme (LDFSS) formulation of Ref. 4, and viscous fluxes are computed using the approach of Ref. 1. Turbulence is modeled using the Shear Stress Transport (SST) model of Ref. 5.

Simulations are evolved to steady-state as determined by a reduction in density and turbulent residuals, as well as the convergence of forces on surfaces in the simulation. The final state is determined from time-averaging of the last 10 seconds of simulation time.

#### 2.1.1 Bus Model

A CAD model of a standard Charlotte Area Transit Systems (CATS) bus, as shown in shown in Fig. 1, was generated based on detailed measurements performed by the authors, with the assistance of CATS operators and management. The CAD model was then verified against the Gillig low-floor bus model layout obtained from CATS through an NDA between UNC Charlotte and CATS; bus HVAC operating condition data were also obtained through the same NDA. Note that Gillig low-bus bus model is the most common variant of CATS buses. This bus is approximately $11.6 \times 2.4 \times 2.5$ m (L×W×H). There are 25 seats for passengers in the main cabin, and the seating at the rear of the bus is elevated.

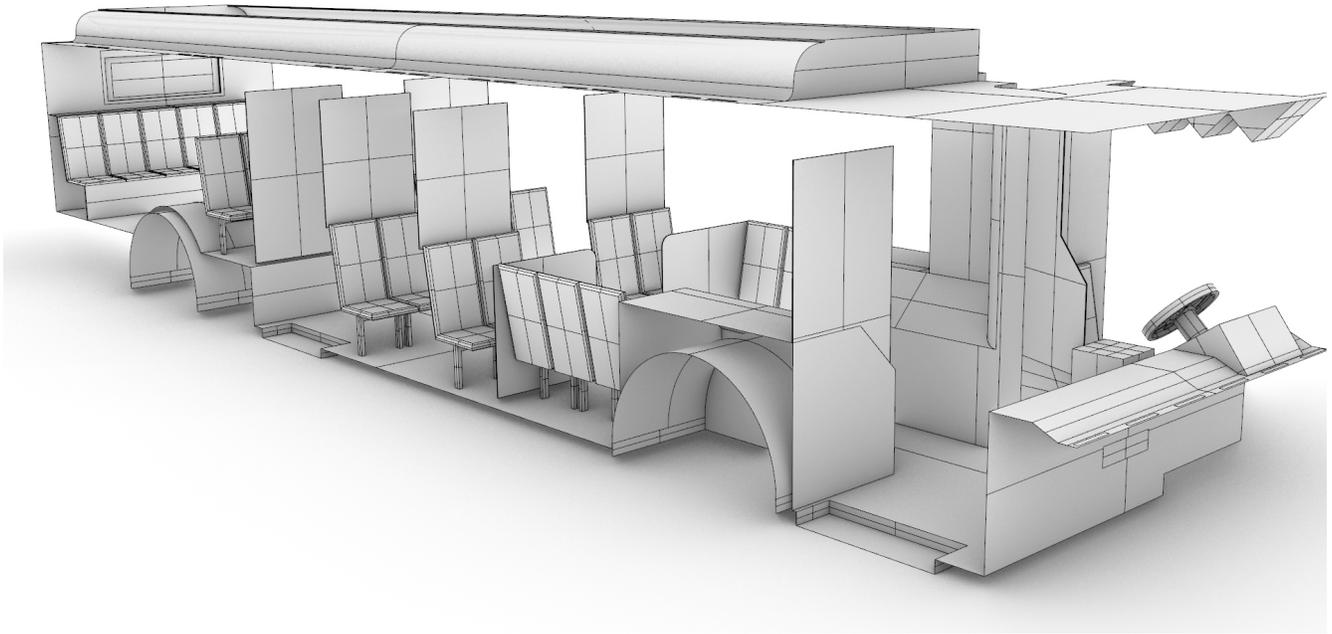

(a) Starboard Side View

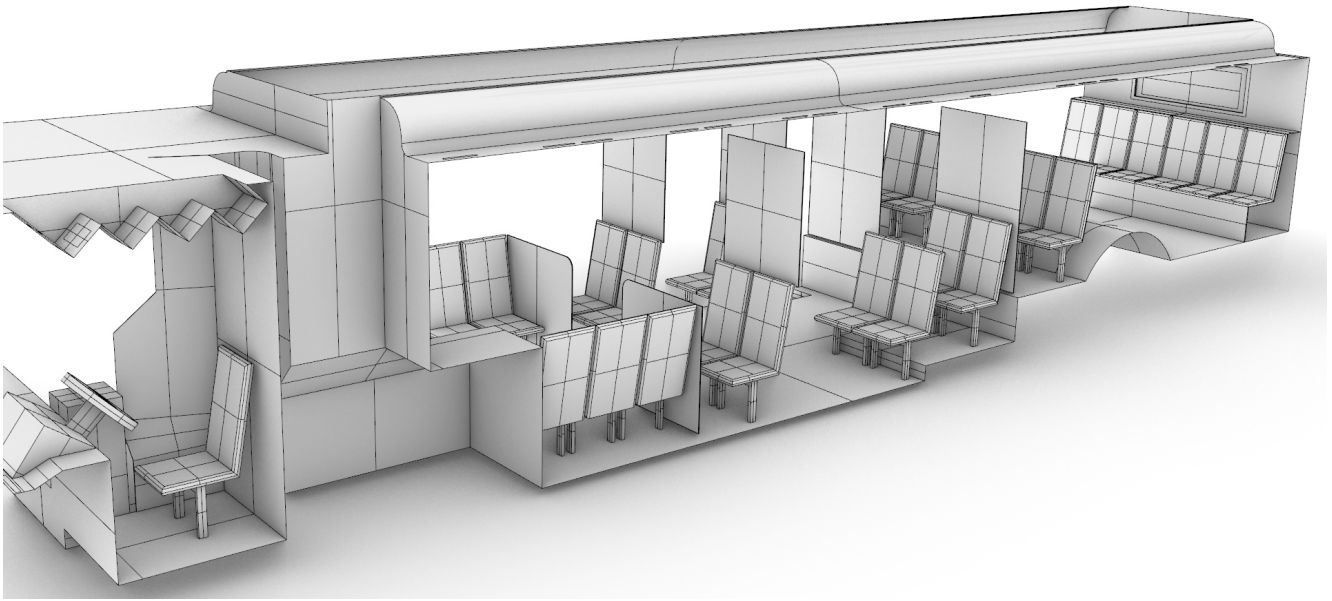

(b) Port Side View

**Figure 1.** CAD model of the CATS bus generated based on measurements by the authors.

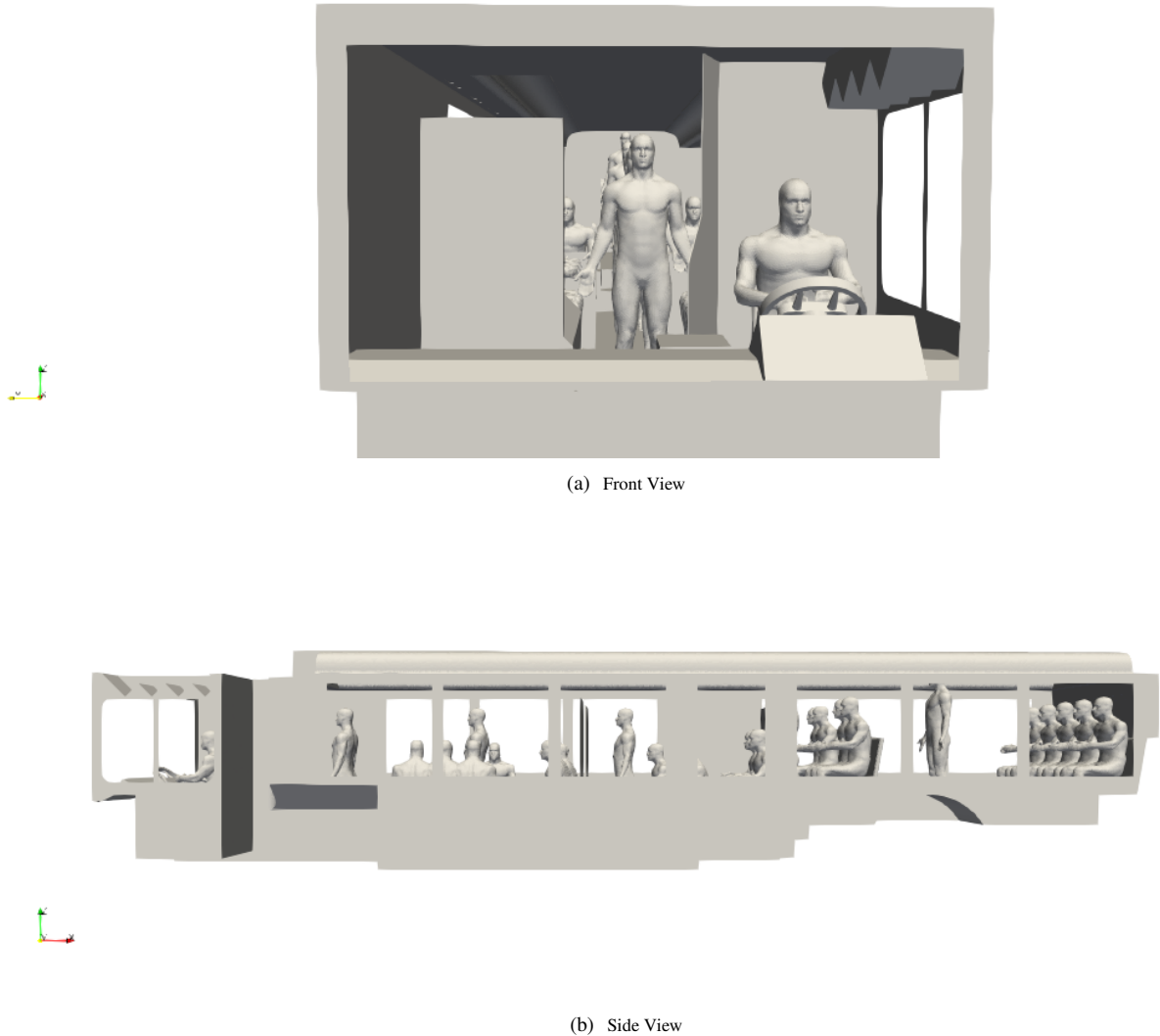

(a) Front View

(b) Side View

**Figure 2.** Integrated bus and passenger models for a bus at maximum occupancy.

### *2.1.2 Passenger Model*
CAD models of both seated and standing passengers were created using Corvid Technologies' proprietary CAVEMAN model, which has previously been used for combat damage mitigation simulations. The passenger models were integrated with the bus model for the various seating and occupancy scenarios described in Sec. 2.2.2. Fig. 2 shows an example of an integrated grid at maximum capacity.

### *2.1.3 Grid Generation*
Conformal grids were created for each bus/passenger combination using a commercial software ANSA by Beta CAE. All surfaces were meshed with triangular elements, including the passenger models (Fig. 3). Prism layers were grown from the surfaces to capture near-wall gradients, and tetrahedral elements were used in the remaining volumes. Inside the bus the maximum cell size (surface and volume) was 25 mm, with additional refinement around curvature and regions of interest. On passengers the maximum surface cell size was 15 mm, with cell sizes of 5 mm near noses and mouths. Volume mesh refinement regions limited volume cell sizes within 100 mm of each passenger to 15 mm.

Initial testing showed that aspect ratio based layer growth with first layer aspect ratio of 0.05 produced a high quality near-wall mesh with average wall y+ values of approximately 0.7. This first layer specification was used for all surfaces in all meshes. Due to the complex nature of the interior flow field, the actual y+ values vary significantly and typically range, on average, from 0.3-1.0 for all interior surfaces. It should be noted that for some of the higher-speed flow configurations y+ can exceed 1.0 in the vicinity of source boundary conditions (see Sec. 2.1.5).

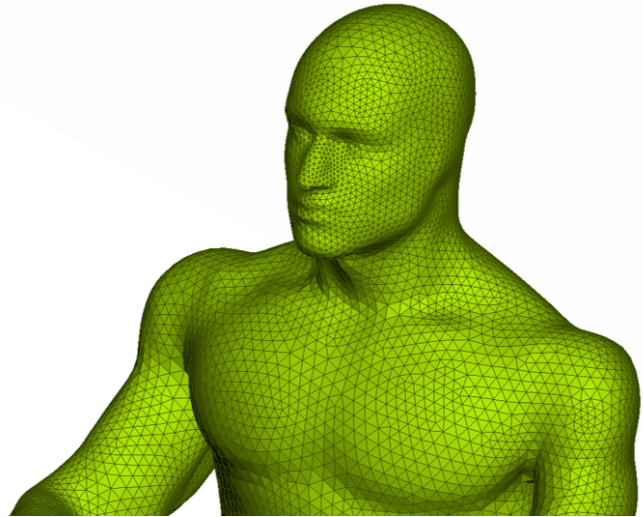

**Figure 3.** Example of triangular surface mesh on a bus passenger model generated using ANSA.

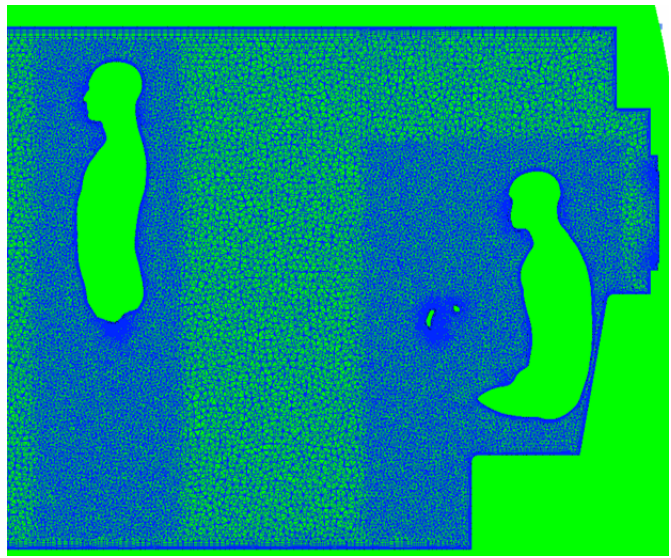

**Figure 4.** Slice through the volume mesh generated using ANSA. Prism layers are visible extending from surfaces, including occupants. Additional refinement is also included around occupants.

**Table 1.** Summary of ambient conditions for the flow solver.

| Variable | Definition | Value |
|---|---|---|
| $\gamma$ | Ratio of specific heats | 1.4 |
| $R$ | Specific gas constant | 287.1 J/(kg K) |
| $T_{f,0}$ | Initial fluid temperature | 288.15 K |
| $P_{f,0}$ | Initial fluid pressure | 101325 Pa |
| $\rho_{f,0}$ | Initial fluid density | 1.225 kg/m$^3$ |
| $\mu_{f,0}$ | Initial fluid dynamic viscosity | $1.79 \times 10^{-5}$ kg/(m s) |

### 2.1.4 Ambient Conditions

The ambient fluid conditions in the bus are treated as air and initialized to a uniform pressure of $P_{f,0} = 101325$ Pa $= 1$ atm and $T_{f,0} = 288.15$ K $= 59^o$F. We assume the air is a calorically perfect ideal gas with ratio of specific heats $\gamma = 1.4$ and specific gas constant $R = 287.1$ J/(kg K). This yields a uniform density of $\rho_{f,0} = 1.225$ kg/m$^3$.

The dynamic viscosity $\mu_{f,0}$ is computed using Sutherland's law[6], defined as

$$\mu_f = \frac{C_1 T^{3/2}}{T+S} \qquad (1)$$

where $C_1$ and $S$ are constants, defined for air as $C_1 = 1.458 \times 10^{-6}$ kg/(m s $\sqrt{K}$) and $S = 110.4$ K. At the initial bus temperature of $T_{f,0} = 288.15$ K, this yields $\mu_{f,0} = 1.79 \times 10^{-5}$ kg/(m s).

Table 1 summarizes the ambient conditions for the flow solver.

### 2.1.5 Boundary Conditions

All surfaces, such as the bus walls, seats, closed windows, and passengers, are treated in the flow solver using an adiabatic no-slip boundary condition. The HVAC registers are modeled as mass flow inlets, which enforce a mass flow rate at a given temperature. The static pressure is extrapolated from the interior of the domain and the fluxes through the boundary are specified directly. Similarly, the HVAC returns are modeled as mass flow outlets. The density, pressure, and velocity are all calculated to provide the specified mass flow rate.

## 2.2 Flow Considerations

The transport of respiratory particles containing SARS-CoV-2 viruses from an infected passenger will be governed by the flow field in the bus. The current study examines the effect of both environmental effects, such as HVAC settings and open windows, as well as mandated COVID-19 interventions by CATS management, such as seating and occupancy restrictions.

### 2.2.1 HVAC System

The transport of respiratory particles will be governed by the flow field in the bus. When the windows and doors are shut, the internal flow dynamics will be largely determined by the HVAC system. The main cabin HVAC system has multiple registers in the cabin ceiling and a single return at the rear of the bus. The main cabin registers blow air downward from the ceiling. The main HVAC system also feeds a driver cabin register, which also blows downward from the ceiling and can be controlled independently from the main cabin. The front of the bus also has defroster HVAC; the intake is in the driver cabin, and the air is exhausted upward on the front window interior. Four settings are considered here: off, low, medium, and high. The locations of the HVAC system components are show in Fig. 5.

Certain components can be adjusted to increase or decrease the airflow rate inside the bus. The main cabin airflow rate is controlled indirectly by specifying a desired temperature. The airflow rate then adjusts automatically. Five flow rates are considered here for the main cabin registers, based on the maximum possible flow rate: off, 1/4 maximum, 1/2 maximum, 3/4 maximum, and maximum. The driver register can be adjusted independently and has three settings: off, low, and high. Finally, the front defroster has four settings: off, low, medium, and high. To characterize the airflow from the HVAC system inside the CATS bus, air speed measurements were taken by the authors using an anemometer. Up to fourteen measurements were performed and averaged for each component. The volume of air contained on the bus is roughly 55 m$^3$. The HVAC system is capable of up to 140 air changes per hour; i.e, the average parcel of air on the bus will be replaced every 26 seconds. Results of our measurements are summarized in Table 2.

### 2.2.2 Seating and Occupancy

In response to the COVID-19 pandemic, CATS has mandated limited seating on buses. To examine the efficacy of these restrictions, we examine multiple seating and occupancy scenarios. We consider passenger arrangements both without and with the current COVID-19 restrictions, as indicated in Fig. 6. CATS currently allows passengers to stand, even when all available seats are in use. We therefore consider three occupancy configurations: maximum seated and standing occupancy, maximum seated occupancy (no passengers standing), and half-seated occupancy (no passengers standing). These configurations are also indicated in Fig. 6.

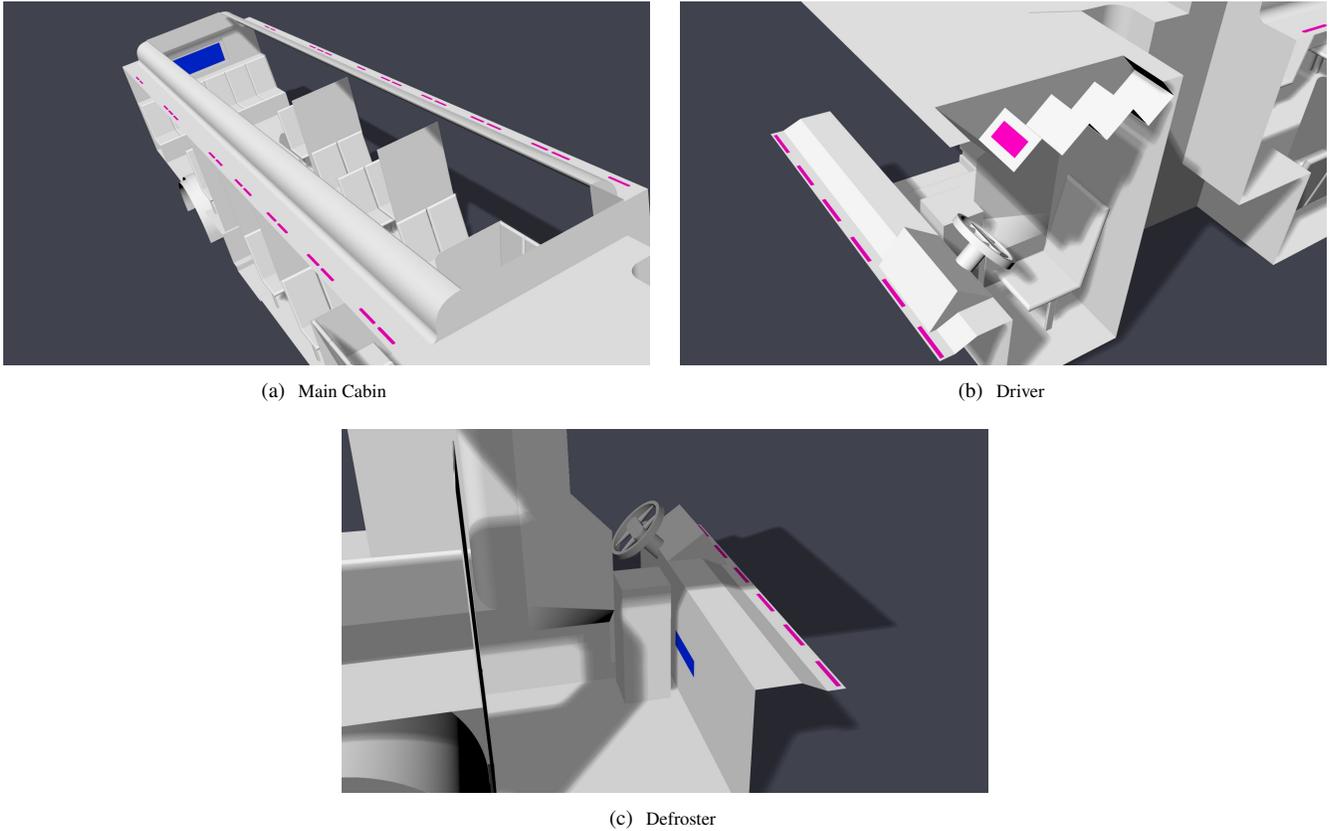

(a) Main Cabin

(b) Driver

(c) Defroster

**Figure 5.** Location of modeled HVAC components. Registers are colored pink, and returns are colored blue.

**Table 2.** Summary of HVAC measurements on a CATS bus. Reported values are averages of up to fourteen measurements.

| HVAC Component | Area (m$^2$) | Setting | Airspeed (m/s) |
|---|---|---|---|
| Main cabin register | 0.08495 | Max | 4.125 |
| Main cabin return | 0.37016 | N/A | 3.0 |
| Driver area register | 0.02065 | Low | 10.0 |
| | | High | 20.0 |
| Defroster (blower) | 0.03871 | Low | 5.0 |
| | | Med | 6.66 |
| | | High | 8.0 |
| Defroster (return) | 0.04935 | Low | 4.0 |
| | | Med | 5.3 |
| | | High | 6.4 |

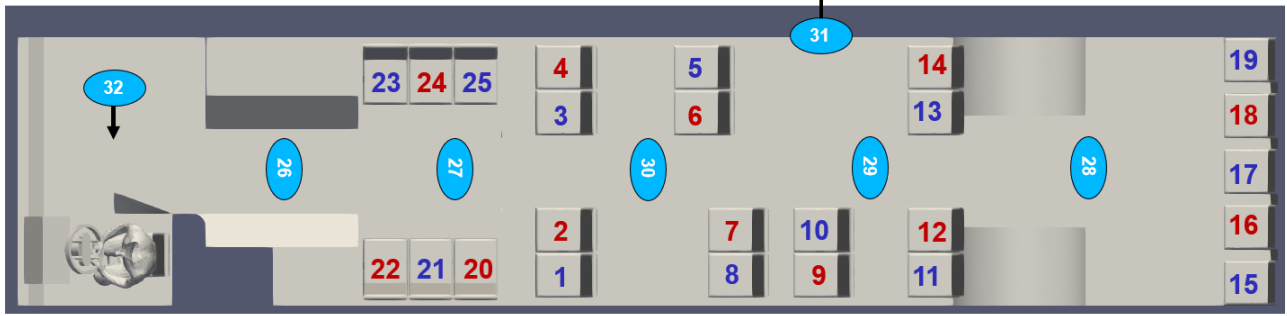

(a) Without COVID-19 Restrictions

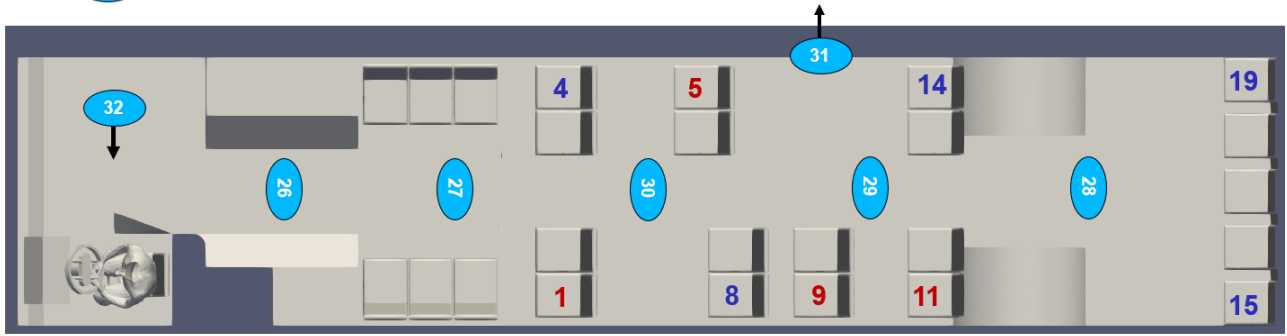

(b) With COVID-19 Restrictions

**Figure 6.** Map of bus interior showing the location and numbering convention of passengers a) without and b) with the current COVID-19 occupancy restrictions. Note that passengers 31 and 32 are only included in simulations in which the doors are open. In the half-seated configuration, only the blue seated passengers are present.

### 2.2.3 Ventilation

Depending on the weather conditions, CATS can open some or all of the windows on the bus. We consider several scenarios: all windows fully shut, all windows fully open, half of the windows fully open, and two mixed configurations labeled as alternate open A and alternate open B.

When the windows are open, the speed of the bus will alter the flow dynamics. We consider two typical speeds of 25 MPH and 35 MPH. The CATS bus considered for this study also has two doors, a forward and an aft door, that open at the same time. We consider a scenario in which the doors are open but the bus is not moving. In this case, the wind in the environment around the bus can also affect the internal flow dynamics. We consider two wind speeds of 6 MPH and 12 MPH, based on the average annual wind speed of roughly 8 MPH in Charlotte, NC[7].

### 2.2.4 Run Groups

Due to the large number of variables considered, the flow simulations are divided into three run groups, labeled A, B, and C, based on the window and door configurations. Run groups A, B, and C are summarized in Tables 3, 4, and 5, respectively.

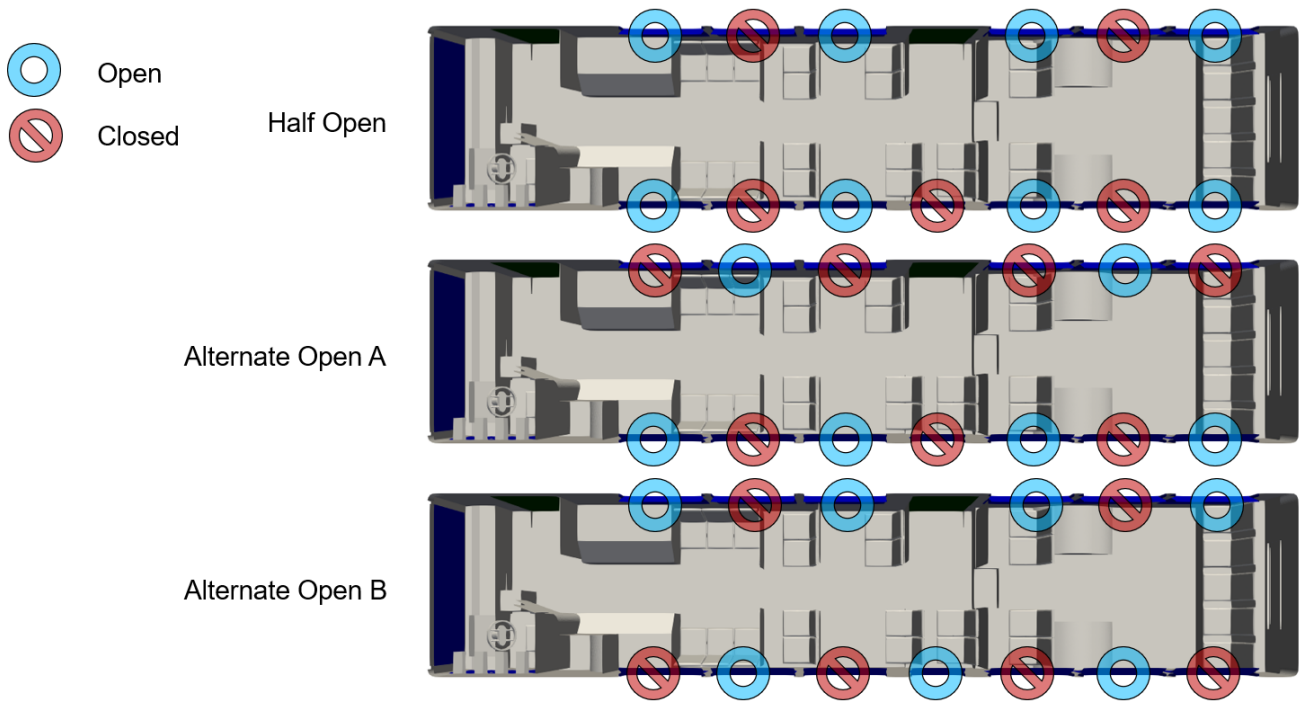

**Figure 7.** Map of bus interior showing the location and configuration of windows.

**Table 3.** Summary of run group A, which is performed with all windows and the door closed. A total of 144 flow simulations are performed for group A.

| Variable | Values |
| --- | --- |
| Windows | all closed |
| Doors | closed |
| Bus Speed | N/A |
| Wind Speed | N/A |
| Main cabin HVAC setting | 1/4 max, 1/2 max, 3/4 max, max |
| Driver area HVAC setting | low, high |
| Defroster setting | low, med, high |
| Seating restrictions | none, current |
| Passenger count | max (standing), max (sitting), half |

**Table 4.** Summary of run group B, which is performed with some or all windows open and the door closed. A total of 48 flow simulations are performed for group B.

| Variable | Values |
| --- | --- |
| Windows | all open, half open, config A, config B |
| Doors | closed |
| Bus Speed | 25 MPH, 35 MPH |
| Wind Speed | 0 MPH |
| Main cabin HVAC setting | off |
| Driver area HVAC setting | off |
| Defroster setting | off |
| Seating restrictions | none, current |
| Passenger count | max (standing), max (sitting), half |

**Table 5.** Summary of run group C, which is performed with all windows closed and the door open. A total of 96 flow simulations are performed for group C.

| Variable | Values |
|---|---|
| Windows | all closed |
| Doors | open |
| Bus Speed | 0 MPH |
| Wind Speed | 6 MPH, 12 MPH |
| Main cabin HVAC setting | 1/4 max, 1/2 max, 3/4 max, max |
| Driver area HVAC setting | low, high |
| Defroster setting | low, med |
| Seating restrictions | none, current |
| Passenger count | max (standing), half |

**2.3 Lagrangian Particle Tracking**

A Lagrangian Particle Tracking (LPT) approach is used to model the transport of SARS-CoV-2 viruses in respiratory droplets. Three injection scenarios are examined for each potential infected bus occupant: speaking, coughing, and sneezing. The injection of particles is modeled as an instantaneous event. The flow conditions are derived from the final time-averaged output of the flow solver.

*2.3.1 Particle Dynamics*

The trajectories of the particles are evolved using the LPT algorithms in VTK[8], an open-source post-processing and visualization tool. The VTK LPT algorithm uses the formulation of Ref. [9]:

$$\frac{d\vec{x}_p}{dt} = \vec{u}_p \quad (2)$$

$$\frac{d\vec{u}_p}{dt} = \frac{\alpha(\vec{u}_f - \vec{u}_p)}{\tau_p} - \vec{g}(1.0 - \frac{\rho_f}{\rho_p}) \quad (3)$$

where $\vec{x}_p$ is the particle location vector, $\vec{u}_p$ is the particle velocity vector, $\alpha$ is a drag coefficient, $\tau_p$ is the particle stopping time, $\vec{u}_f$ is the fluid velocity vector, $\rho_f$ is the fluid density, $\rho_p$ is the density of the particle, which we assume is constant at roughly the density of water such that $\rho_p = 1000\,\mathrm{kg/m^3}$, and $\vec{g} = [0, 0, 9.8]\,\mathrm{m/s^2}$ is the gravitational acceleration vector.

The drag coefficient $\alpha$ is defined as

$$\alpha = 1.0 + 0.15\,\mathrm{Re}_p^{0.687} \quad (4)$$

with $\mathrm{Re}_p$ the Reynolds number of the particle defined as

$$\mathrm{Re}_p = \frac{\rho_f\,|\vec{u}_f - \vec{u}_p|\,d_p}{\mu_f} \quad (5)$$

where $d_p$ is the particle diameter and $\mu_f$ is the dynamic viscosity of the fluid, computed using Sutherland's Law (Eq. 1).

The particle stopping time $\tau_p$ is defined as

$$\tau_p = \frac{\rho_p\,d_p^2}{18\,\mu_f}. \quad (6)$$

The trajectories are integrated forward in time using an adaptive fourth order Runge-Kutta algorithm with fifth order correction (RK45). The necessary fluid quantities (density $\rho_f$, velocity vector $\vec{u}_f$, and dynamic viscosity $\mu_f$) are interpolated from the flow solver output to the particle location at each particle time step.

The stopping time $\tau_p$ is a characteristic time over which the relative velocity between the particles and the fluid is brought to zero. In other words, after one or more stopping times, the particle can be assumed to move with the fluid. Fig. 8 shows the stopping time as a function of the initial particle diameter at the initial bus flow conditions.

The stopping time is directly proportional to the square of the particle diameter; therefore the smallest particles more rapidly become entrained in the fluid flow. Due to the computational cost associated with modeling the smallest particles using the LPT dynamics, which would limit the timestep to approximately the stopping time, we instead model the smallest particles using the VTK "Streamline" integrator. The streamline integration is similar to the LPT algorithm; however, it is much simpler, as the particle velocity is assumed to be the local fluid velocity, and no drag equation need be solved. Note that this approach neglects gravity, which is a reasonable assumption for the small particles where the gravitational settling time is on the order of minutes or hours, much longer than the HVAC air change time (roughly 30 seconds).

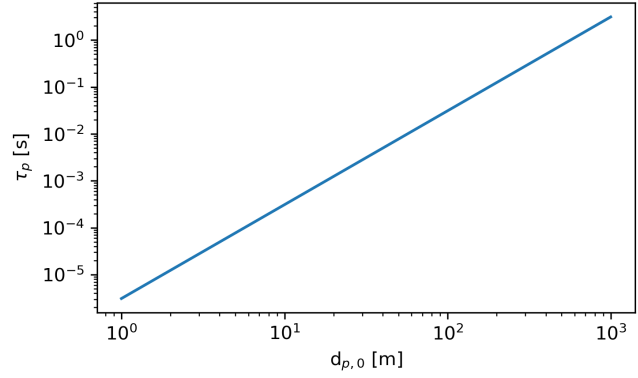

**Figure 8.** The stopping time $\tau_p$ as a function of initial particle diameter $d_{p,0}$. The particle density is constant at $\rho_p = 1000\,\text{kg/m}^3$. The fluid dynamic viscosity $\mu_f$ is computed using Sutherland's Law (see Eq. 1) at the initial bus temperature $T_{f,0} = 288.15\,\text{K}$.

We have verified the accuracy of the use of streamlines by comparing results to LPT trajectories of particles. For particles with initial diameter $d_{p,0} \leq 25\,\mu\text{m}$, the results obtained with the LPT algorithm and the streamline algorithm are indistinguishable; however, the streamline method required several orders of magnitude fewer iterations, greatly reducing the computational expense. However, for particles with initial diameter greater than 25 $\mu$m, the deviations between the LPT and the streamline integration are apparent. We therefore use the streamline integration for particles $d_{p,0} \leq 25\,\mu\text{m}$, and and the LPT algorithm for all larger particles in the size distribution.

The LPT formulation does not include Brownian motion. In a similar study of SARS-CoV-2 particle transport, Ref. 10 estimated that the typical deviation of a respiratory particle due to Brownian motion is approximately 0.03 m in 5000 seconds, which is negligible compared to the total distance traveled.

### 2.3.2 Cunningham Slip Correction Factor

Very small particles ($d < 1\mu$) require a correction to the drag force known as the Cunningham slip correction factor (e.g., Ref 11):

$$C_C = 1.0 + \text{Kn}(2.514 + 0.80 \exp\frac{-0.55}{\text{Kn}}) \tag{7}$$

where $\text{Kn} = \frac{\lambda}{d_p}$ is the Knudsen number, $d_p$ is the particle diameter, and $\lambda$ is the mean free path of the gas (air in our simulation). We neglect small deviations in temperature and pressure in the domain and set the mean free path to be a constant value based on the initial pressure $P_0 = 101325\,\text{Pa}$ and temperature $T_0 = 288.15\,\text{K}$:

$$\lambda_0 = \frac{k_B T_0}{\sqrt{2}\pi d_m^2 P_0} \approx 68\,\mu\text{m} \tag{8}$$

with the average diameter of an air molecule $d_m \approx 0.36\,\text{nm}$. We have modified the LPT algorithms in VTK to include this factor in the Matida drag law. Note that for large particles, the correction factor is still included but is essentially unity.

### 2.3.3 Particle Evaporation

The respiratory droplets contain a large amount of volatiles that will evaporate in the air. Evaporation will reduce the effective diameter of the particle and alter the particle dynamics. We have therefore modified the LPT algorithms in VTK to include particle evaporation.

A time-dependent equation for the diameter of respiratory particles is given by Ref. 12:

$$d_p(t) = (d_{p,0}^2 - 8 v_m D_{\text{H}_2\text{O}} \frac{(P_{\text{sat}} - P_{\text{H}_2\text{O}})}{k_B T})^{1/2} \tag{9}$$

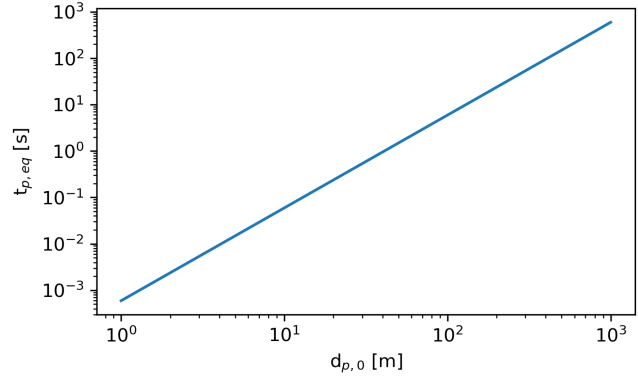

**Figure 9.** The time required to reach the equilibrium diameter, $d(t_{p,eq}) = d_{p,eq} = 0.5\,d_{p,0}$, as a function of initial particle diameter $d_{p,0}$.

where $d_{p,0}$ is the initial diameter of the particle, $v_m = 3 \times 10^{-29}$ m$^3$ is the condensed-phase volume occupied by a single water molecule, $D_{H_2O} = 1.8 \times 10^{-5}$ m$^2$/s is the molecular diffusivity of water vapor in air, $P_{sat}$ is the partial pressure of water vapor in equilibrium with the particle surface, $P_{H_2O}$ is the partial pressure of water vapor in ambient air, $T$ is the temperature, and $k_b = 1.38 \times 10^{-23}$ J/K is Boltzmann's constant. $P_{H_2O}$ is equivalent to $P_{sat}$ multiplied by the relative humidity (RH), $P_{H_2O} = (RH)\,P_{sat}$. We treat both the relative humidity and the saturation pressure as constant at RH = 0.50 and $P_{sat} = 2320$ Pa.

Following Ref. 12, we approximate the equilibrium diameter of a particle $d_{p,eq}$ to be roughly half of its initial diameter;

$$d_{p,\text{eq}} \approx 0.5 d_{p,0}. \tag{10}$$

Fig. 9 shows the time required to reach the equilibrium diameter, $t_{p,eq}$, as a function of initial diameter. The smallest particles rapidly evaporate to their equilibrium diameter, while the largest particles require substantially more time.

### 2.3.4 Particle Distributions

Biological particles of saliva and mucus are introduced into the air by means of typical respiratory functions: speaking, coughing, and sneezing. A wide and varying distribution of respiratory particle sizes are generated in each of these events. Although there is substantial variation in the literature on the exact size distribution of the particles, likely due to differences in sample collection, most particles are typically within the range of 1 — 1000 $\mu$m[10,11,13].

Contraction of viruses can occur through breathing in small particles ($\lesssim 10$ $\mu$m) as well as touching contaminated surfaces[12]. This provides motivation to understand the paths of particles with a wide range of diameters to determine ways to reduce the chance of virus contraction.

We independently model three possible injection events: speaking, a cough, and a sneeze. Each event is treated as an instantaneous injection of a given particle size distribution. For speaking and a cough, we use the distributions reported by Ref. 14 sampled at a distance of 10 mm. For a sneeze, we use the distribution reported by Ref. 15. For consistency, we resample these distributions to a single distribution given in Table 6. Fig. 10 compares the cumulative distributions of the references with the resampled distributions used in this study. Fig. 11 compares the number of particles in each diameter sampling interval between the references and this study.

Table 7 summarizes the particle injection speeds and diameter distributions.

### 2.3.5 Viral Mean Exposure Time

To enable comparison of the numerous runs and infection mitigation strategies, the relative amount of exposure to viruses is quantified using a volumetric mean exposure time (MET)[16–18]. This quantity is similar to a particle residence time (PRT) and is defined over a discrete element.

We define the viral MET (VMET) at a discrete volume element $e$ as

$$\text{VMET}(e) = \frac{1}{(V_e)^{(1/3)}} \sum_{p=1}^{N_p} N_v(p) \int_{t=0}^{\infty} \chi_e(p,t)\,dt \tag{11}$$

where $V_e$ is the volume of element $e$, $N_p$ is the total number of particles injected, $\chi_e(p,t)$ is a sharp Heaviside function defined as unity if particle $p$ resides in the volume $V_e$ at time $t$ and zero otherwise, and $N_v(p)$ is the estimated number of SARS-CoV-2 viruses in particle $p$.

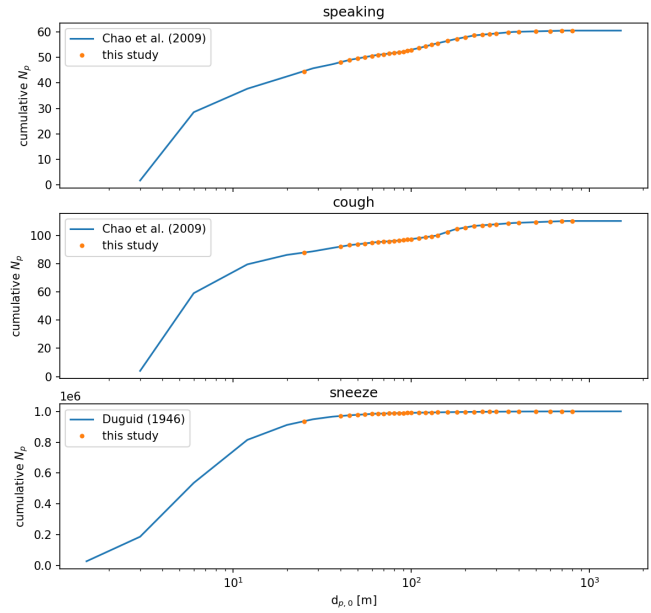

**Figure 10.** Comparison of cumulative particle number distributions between the source data in Refs. 14 and 15 and the resampled size distributions for this study.

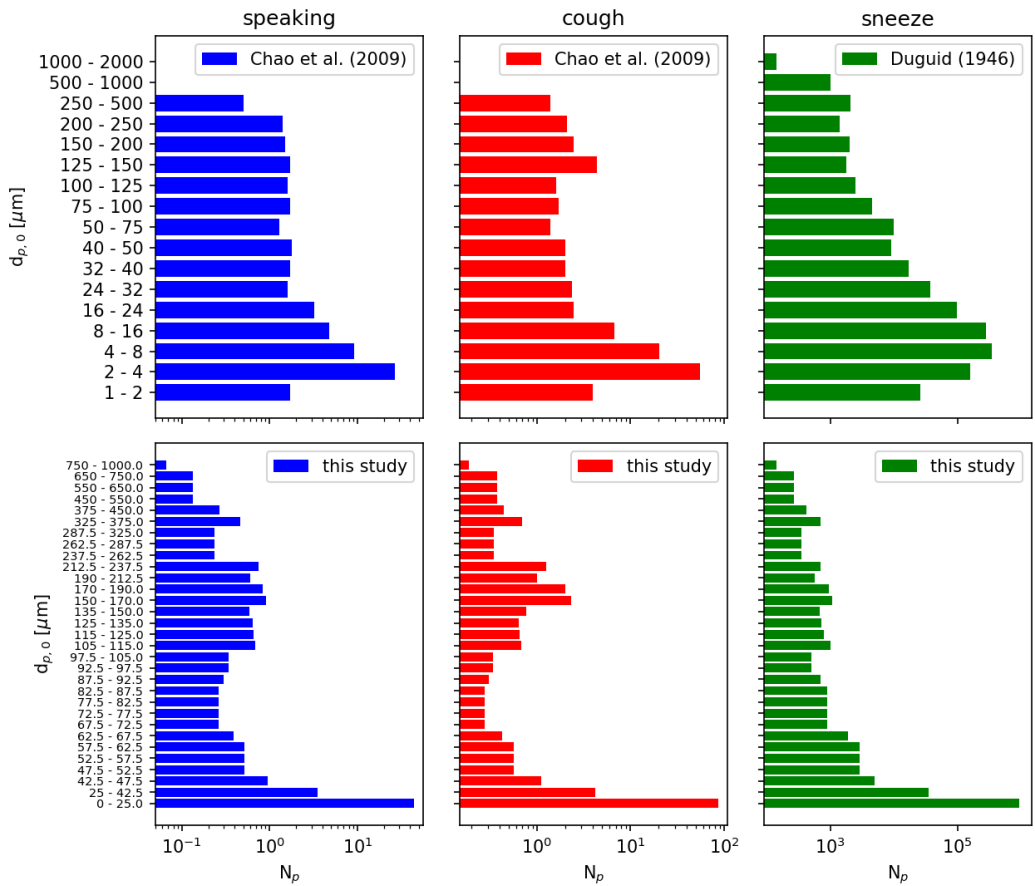

**Figure 11.** Particle number distributions of initial particle diameter $d_{p,0}$. The data from Refs. 14 and 15 is resampled to a finer particle size distribution for this study. Note that the number of particles reported is a sum over each sampled diameter range.

**Table 6.** Particle initial diameter sampling distribution. The average number of particles injected $N_p$ and the average number of copies of the SARS-CoV-2 virus injected $N_v$ are also shown.

| Diameter Range [$\mu$m] | LPT Sample Diameter [$\mu$m] | Average $N_p$ (speaking) | Average $N_v$ (speaking) | Average $N_p$ (cough) | Average $N_v$ (cough) | Average $N_p$ (sneeze) | Average $N_v$ (sneeze) |
|---|---|---|---|---|---|---|---|
| 0.0 – 25.0 | streamlines | 44.50 | 2.55 | 87.66 | 5.02 | 936247.16 | 53617.58 |
| 25.0 – 42.5 | 40 | 3.56 | 0.83 | 4.23 | 0.99 | 34879.55 | 8181.77 |
| 42.5 – 47.5 | 45 | 0.94 | 0.32 | 1.11 | 0.37 | 5000.65 | 1670.17 |
| 47.5 – 52.5 | 50 | 0.51 | 0.24 | 0.57 | 0.26 | 2857.52 | 1309.17 |
| 52.5 – 57.5 | 55 | 0.51 | 0.31 | 0.57 | 0.35 | 2857.52 | 1742.50 |
| 57.5 – 62.5 | 60 | 0.51 | 0.41 | 0.57 | 0.45 | 2857.52 | 2262.24 |
| 62.5 – 67.5 | 65 | 0.39 | 0.39 | 0.43 | 0.43 | 1878.82 | 1891.13 |
| 67.5 – 72.5 | 70 | 0.26 | 0.33 | 0.28 | 0.35 | 900.12 | 1131.59 |
| 72.5 – 77.5 | 75 | 0.26 | 0.40 | 0.28 | 0.43 | 900.12 | 1391.81 |
| 77.5 – 82.5 | 80 | 0.26 | 0.49 | 0.28 | 0.53 | 900.12 | 1689.14 |
| 82.5 – 87.5 | 85 | 0.26 | 0.59 | 0.28 | 0.63 | 900.12 | 2026.06 |
| 87.5 – 92.5 | 90 | 0.30 | 0.80 | 0.31 | 0.83 | 700.09 | 1870.59 |
| 92.5 – 97.5 | 95 | 0.34 | 1.07 | 0.34 | 1.07 | 500.07 | 1571.43 |
| 97.5 – 105.0 | 100 | 0.34 | 1.25 | 0.34 | 1.25 | 500.07 | 1832.83 |
| 105.0 – 115.0 | 110 | 0.68 | 3.32 | 0.68 | 3.32 | 1000.13 | 4879.01 |
| 115.0 – 125.0 | 120 | 0.65 | 4.12 | 0.65 | 4.12 | 790.10 | 5004.08 |
| 125.0 – 135.0 | 130 | 0.64 | 5.15 | 0.64 | 5.15 | 720.09 | 5798.50 |
| 135.0 – 150.0 | 140 | 0.59 | 5.97 | 0.77 | 7.78 | 673.42 | 6772.79 |
| 150.0 – 170.0 | 160 | 0.91 | 13.61 | 2.35 | 35.23 | 1066.81 | 16015.56 |
| 170.0 – 190.0 | 180 | 0.83 | 17.74 | 2.01 | 42.96 | 940.12 | 20095.49 |
| 190.0 – 212.5 | 200 | 0.60 | 17.59 | 1.00 | 29.32 | 560.07 | 16422.20 |
| 212.5 – 237.5 | 225 | 0.75 | 31.31 | 1.25 | 52.19 | 700.09 | 29227.99 |
| 237.5 – 262.5 | 250 | 0.23 | 13.36 | 0.35 | 20.04 | 350.05 | 20046.63 |
| 262.5 – 287.5 | 275 | 0.23 | 17.79 | 0.35 | 26.68 | 350.05 | 26682.07 |
| 287.5 – 325.0 | 300 | 0.23 | 23.09 | 0.35 | 34.64 | 350.05 | 34640.58 |
| 325.0 – 375.0 | 350 | 0.47 | 73.33 | 0.70 | 110.00 | 700.09 | 110015.91 |
| 375.0 – 450.0 | 400 | 0.27 | 62.55 | 0.44 | 103.99 | 416.72 | 97751.19 |
| 450.0 – 550.0 | 500 | 0.13 | 61.09 | 0.37 | 171.04 | 266.70 | 122188.99 |
| 550.0 – 650.0 | 600 | 0.13 | 105.56 | 0.37 | 295.56 | 266.70 | 211142.58 |
| 650.0 – 750.0 | 700 | 0.13 | 167.62 | 0.37 | 469.34 | 266.70 | 335286.59 |
| 750.0 – 1000.0 | 800 | 0.07 | 125.11 | 0.19 | 350.29 | 142.69 | 267760.07 |
| Total | | 60.50 | 758.27 | 110.10 | 1774.62 | 1001440.00 | 1411918.24 |

**Table 7.** Summary of particle injection event parameters. For the distributions from Ref. 14, we use the samples collated at a distance of 10 mm.

| Event | Distribution Reference | Injection Speed[11] |
|---|---|---|
| Speaking | 14 | 1 m/s |
| Cough | 14 | 10 m/s |
| Sneeze | 15 | 35 m/s |

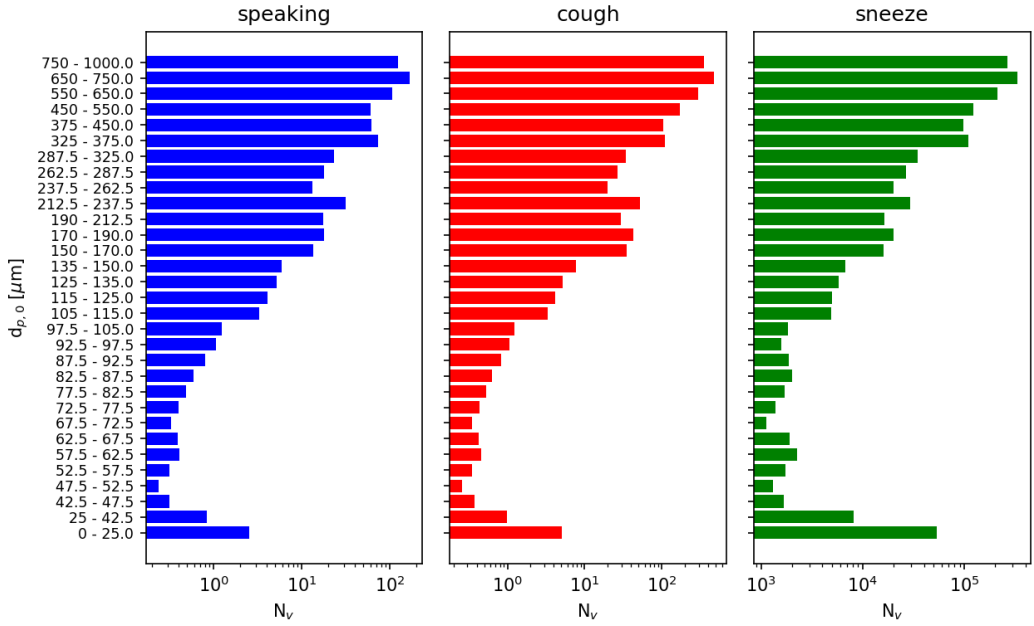

**Figure 12.** Number of copies of SARS-CoV-2 virus injected as a function of initial particle diameter $d_{p,0}$. We assume a viral load of $7 \times 10^{12}$ copies/m$^3$ based on Ref. 19. The number of viruses reported is a sum over each sampled diameter range.

We assume that SARS-CoV-2 viruses are initially uniformly distributed in a given volume of respiratory fluid, at a constant number density of $n_v = 7 \times 10^{12}$ copies/m$^3$[19]. Note that this is an average viral load; the maximum number of copies can be orders of magnitude greater[19]. Furthermore, we assume that the viruses are not evaporated but remain in the particle; therefore the number of viruses in a given particle is purely a function of its initial diameter:

$$N_v(p) = \frac{4}{3} \pi n_v \left(\frac{d_{p,0}}{2}\right)^3 \tag{12}$$

Fig. 12 shows the number of viruses released in each injection event in each diameter range sampled.

We note that our definition of VMET differs from the MET of Ref. 16 in two ways. First, we include the viral load in the definition because we are focused on the risk of transmission of COVID-19. This weights the exposure to a larger particle more heavily than a smaller particle, as the larger particle contains a greater number of copies of the SARS-CoV-2 virus. Second, Ref. 16 included a normalizing factor based on the number of unique 'encounters' of a particle with a volume element. The authors noted that this resulted in recirculating particles contributing less to the MET than a stagnant particle . While this may be true in some circumstances, we do not see a need to include this factor, as our volume elements are quite small and recirculation can still result in exposure.

To compute the VMET, we first resample the particle trajectories to a uniform time step of $\Delta t = 0.001$ s. To obtain consistent volume elements, we compute the VMET on a uniform Cartesian grid of cell length $\Delta x = 0.01$ m. These two steps ensure that no particle will move more than one cell length in a given timestep based on the maximum particle injection speed is 35 m/s. The volumetric VMET computed on the structured grid can then be resampled back to the unstructured grid used in the flow solver for visualization and analysis. The VMET is a cell-centered quantity with units of viral copies multiplied by time per length (e.g., copies-s/m). Intuitively, the VMET represents the average amount of viruses present in a given volume element when sampled over a finite amount of time. The greater the VMET, the higher the risk of exposure to SARS-CoV-2.

We also note that the VMET is not normalized by the total amount of viruses in the injection event. A sneeze produces roughly 10,000 times more respiratory particles than speaking or coughing (and correspondingly 1,000 times more viruses) and is therefore much more dangerous. Our definition of VMET captures this when comparing injection events.

## 3 Validation

### 3.1 Flow Solver Validation

To validate the ability of RavenCFD to simulate low-speed flows, we compared to the results of Ref. 20. The authors presented both experimental and simulation results of cross-ventilation flow through an isolated building. Similar results are reported in

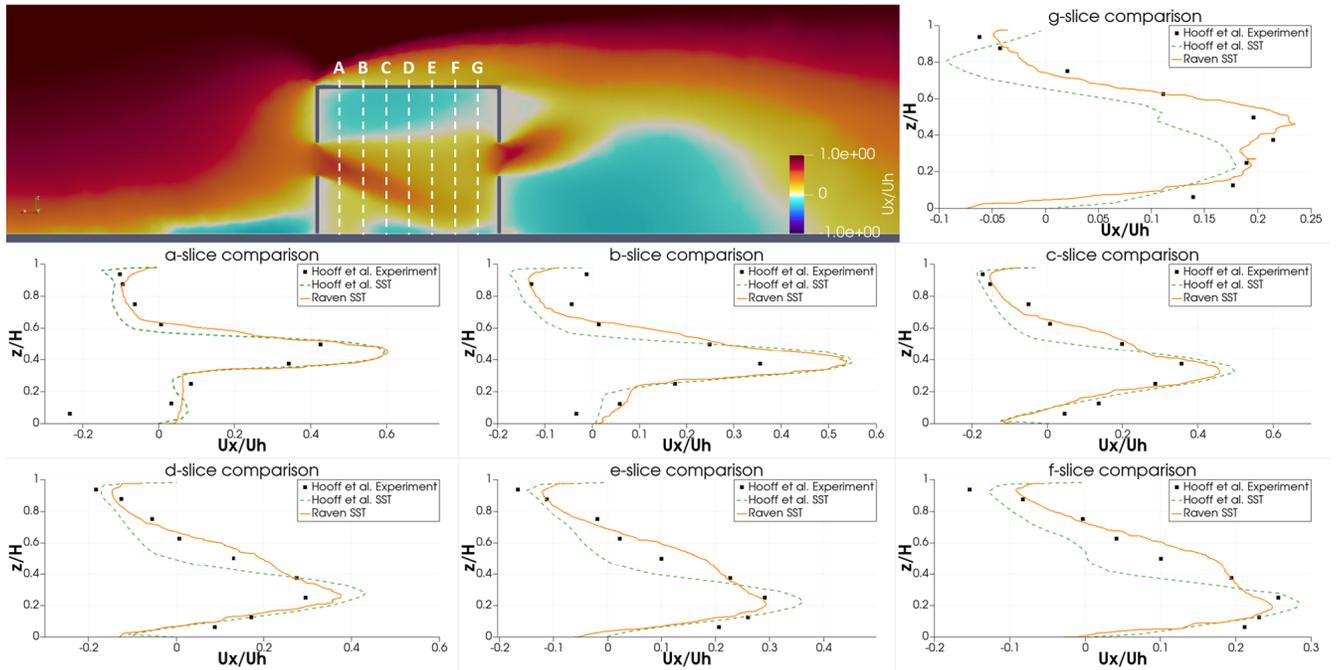

**Figure 13.** Vertical velocity profiles extracted from RavenCFD simulation results at varying position inside the room embedded in cross-flow, compared with the results of Ref. 20.

Refs. 21 and 22.

A building of dimensions $0.2 \times 0.2 \times 0.16\,\mathrm{m}^3$ is placed in a computational domain of extent $2.16 \times 1.8 \times 0.96\,\mathrm{m}^3$. The building has an open window of area $3.3 \times 10^{-3}\,\mathrm{m}^2$ on both the windward and leeward walls.

The velocity is injected into the domain with a vertical power-law profile designed to match the experimental wind tunnel; see eq. 1 of Ref. 20. The velocity is normalized using a reference velocity of $U_H = 4.3\,\mathrm{m/s}$ which is computed at the building height $H = 0.16\,\mathrm{m}$.

RANS simulations are performed in RavenCFD using the SST model of Ref. 5. The turbulent kinetic energy $k$ and specific dissipation rate $\omega$ are also initialized with a vertical profile; see Eqs. 2–5 of Ref. 20.

Simulations are evolved to steady-state, as determined by a reduction of density residuals and by convergence of the forces and moments on the buliding. The results are time-averaged over 7 seconds.

Qualitatively, the authors identify several flowfield features in the RANS simulations of Ref. 20. First, the flow entering the building through the windward window creates a jet that points downward and spreads with a characteristic width. Second, the flow leaving building creates a jet that points upward and also spreads with a characteristic width. Finally, the impinging flow on the top of the building creates a stagnation region. It is worth noting that none of the RANS methods considered in Ref. 20 captured the vertical jet flapping observed in the experiments and LES simulations. Despite this, the authors find that the RANS models do sufficiently capture the volume flow rates and general flow dynamics.

For comparison to experimental results, the horizontal velocity profile is extracted at various locations inside the building. Fig. 13 shows the experimental and SST simulation results of Ref. 20, as well as the RavenCFD SST results. Overall, the RavenCFD results show good agreement with both the experimental measurements and the SST results of Ref. 20, providing high confidence in the ability of RavenCFD to model subsonic cross-flow scenarios similar to those on the CATS bus.

### 3.2 Lagrangian Particle Tracking Validation

The Lagrangian particle tracking algorithm in VTK, including the modifications described above to include the Cunningham slip correction factor and the evaporation model, are validated by comparing to results obtained with an external ODE solver implemented in Python. The external solver uses the "ivp_solve" routine from the "scipy"[23] package.

Particle trajectories are simulated in an empty room with either static or uniform flow. Particles are injected from a known position and with constant velocity. Fig. 14 shows a comparison between the results obtained with the Python code with the VTK LPT results for a single particle injected into static flow. The LPT results from VTK are nearly identical to the results obtained with the external tool, providing high confidence in the ability of VTK to model the particles as well as validating our modifications to the VTK code to include evaporative effects and the Cunningham correction factor.

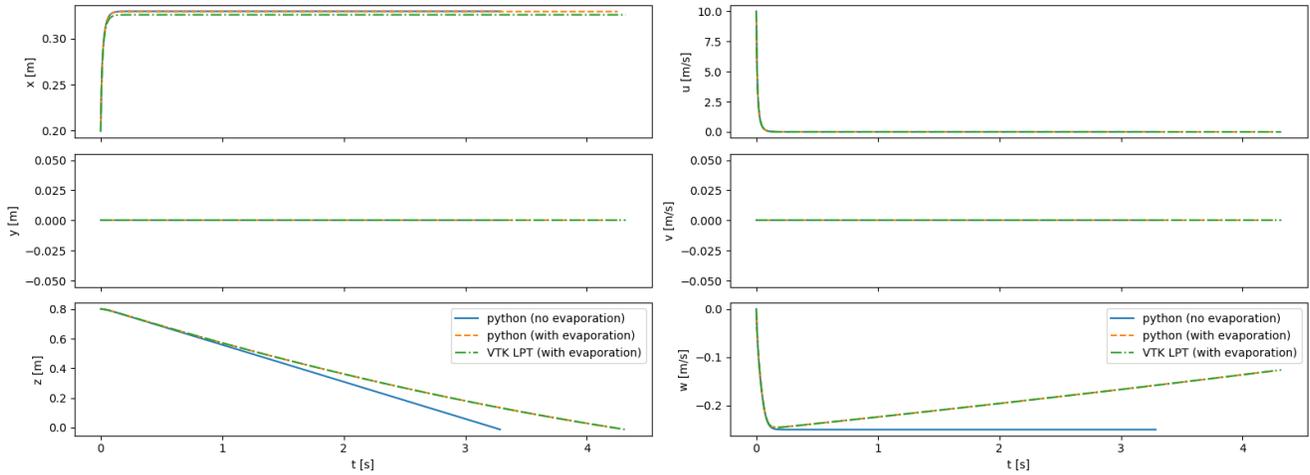

**Figure 14.** Comparison between trajectories computed with Python, both with and without the evaporation model; and with the VTK LPT algorithm. Results are shown for a 100 $\mu m$ particle injected into a static room with initial velocity of 10 m/s in the +x direction. The particles are integrated until they impact the ground.

# 4 Preliminary Results

Due to the large number of cases considered as a part of the current study, a thorough investigation is still ongoing and will be presented in a forthcoming publication(s). In order to provide some perspective on the study results, we share preliminary observations comparing a windows closed case with arbitrarily selected HVAC settings to the windows fully open and half open cases at 35 MPH.

Preliminary results are shown for a fiducial case with windows and doors closed, at half-seated occupancy under COVID-19 restrictions, with the main HVAC at maximum flow rate, the driver HVAC on low, and the defroster on medium. The VMET is presented for injection events from the bus driver to show a comparison of the three types of events: speaking, coughing, and sneezing.

Fig. 15 shows slices of VMET taken through the domain for the fiducial case and centered on the driver. For all injection events (speaking, coughing, and sneezing), the viral load is most concentrated in the area immediately in front of the driver. This is because most of the viral load is contained in the largest particles which have the most inertia and are therefore least affected by the flow dynamics; they are also the most massive and therefore drop rapidly due to gravity.

Fig. 16 shows a volumetric rendering of VMET in the domain which identifies any cells with a non-zero VMET. Although the viral load is concentrated in the area surrounding the driver, there is still an exposure risk from the smaller particles (i.e., aerosols) that are entrained in the flow and can spread throughout the bus. The smallest particles are correspondingly light and do not rapidly drop to the ground. The overall flow dynamics in the bus with the windows and doors closed is directed toward the rear of the bus, which is where the main HVAC return is located. This can be seen in Fig. 17, where the positive velocity in the x-direction indicates flow toward the rear of the bus. Therefore the aerosolized particles will migrate backwards from the driver.

To help determine the best HVAC/window configuration for mitigating the spread of viruses, the previous case is compared to cases with the windows fully open and half open with the HVAC off. The VMET is presented for a sneeze injection event for the driver. Fig. 18 compares the volumetric VMET when bus driver sneezes for each case. For both the windows fully open and half open cases, the VMET for the driver remains largely the same. While the viral load is more concentrated in the area surrounding the driver, there is no flow directed toward the rear of the bus as the AC remains off and the particles are instead ejected from the window just rear of the driver.

While this initially leads to the conclusion that the windows fully/half open cases are better for the prevention of viral transmission, this does not represent the overall flow dynamics in the bus. To further investigate the differences, we consider a sneeze from two arbitrarily selected passengers (14 and 15). Fig. 19 compares the volumetric VMET for each case when passenger 14 sneezes. For the windows closed case, the aerosols emitted by passenger 14 largely move towards the rear of the bus where the main HVAC return is located. In contrast, the windows fully open case shows the aerosols circulating throughout the entire bus. This is a result of the increased turbulence within the bus from having multiple competing inlets in the form of open windows. The windows half open case shows a similar result, only to a slightly lesser degree. Fig. 20 compares the volumetric VMET for each case when passenger 15 sneezes. The results are comparable to those for passenger 14, with the windows closed case directing a majority of the particles aside from the initial injection into the HVAC return. Both cases with windows open similarly show an increase in circulation of aerosols, with the windows fully open case again circulating more of the particles with a higher VMET.

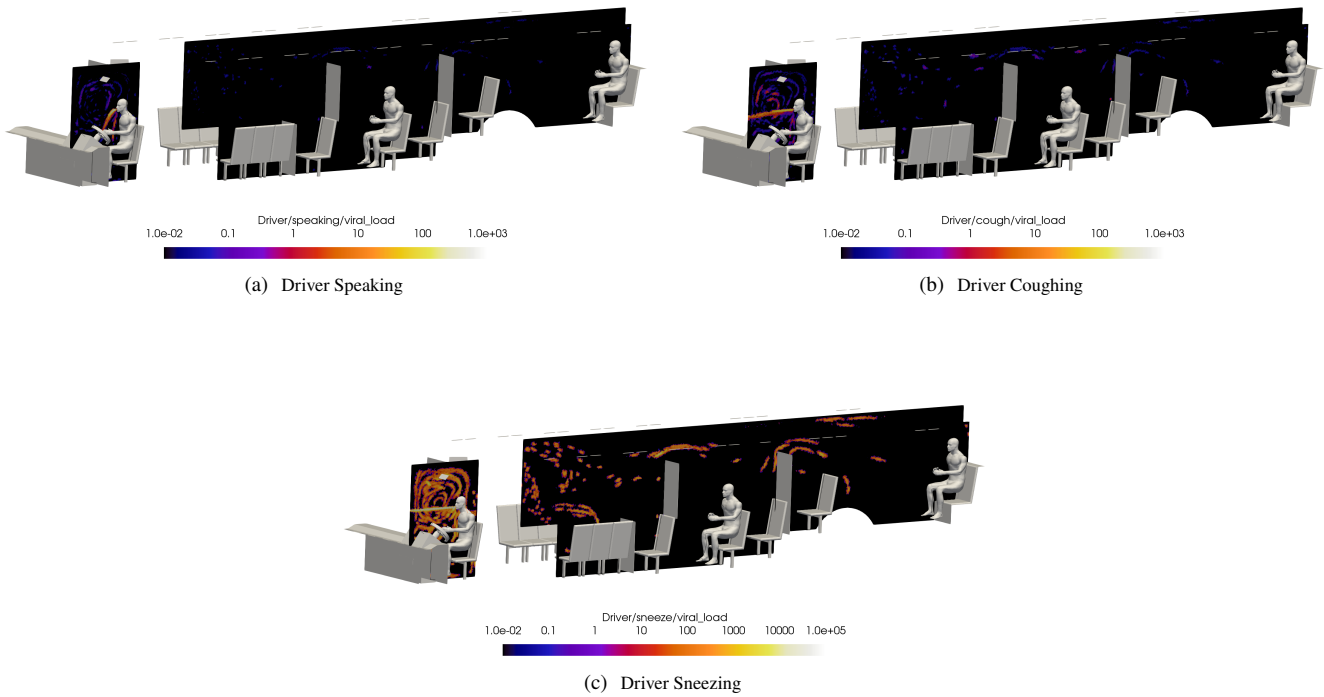

**Figure 15.** Slices of VMET along the bus and centered on the driver for the fiducial case, comparing three injection events of the bus driver a) speaking, b) coughing, and c) sneezing.

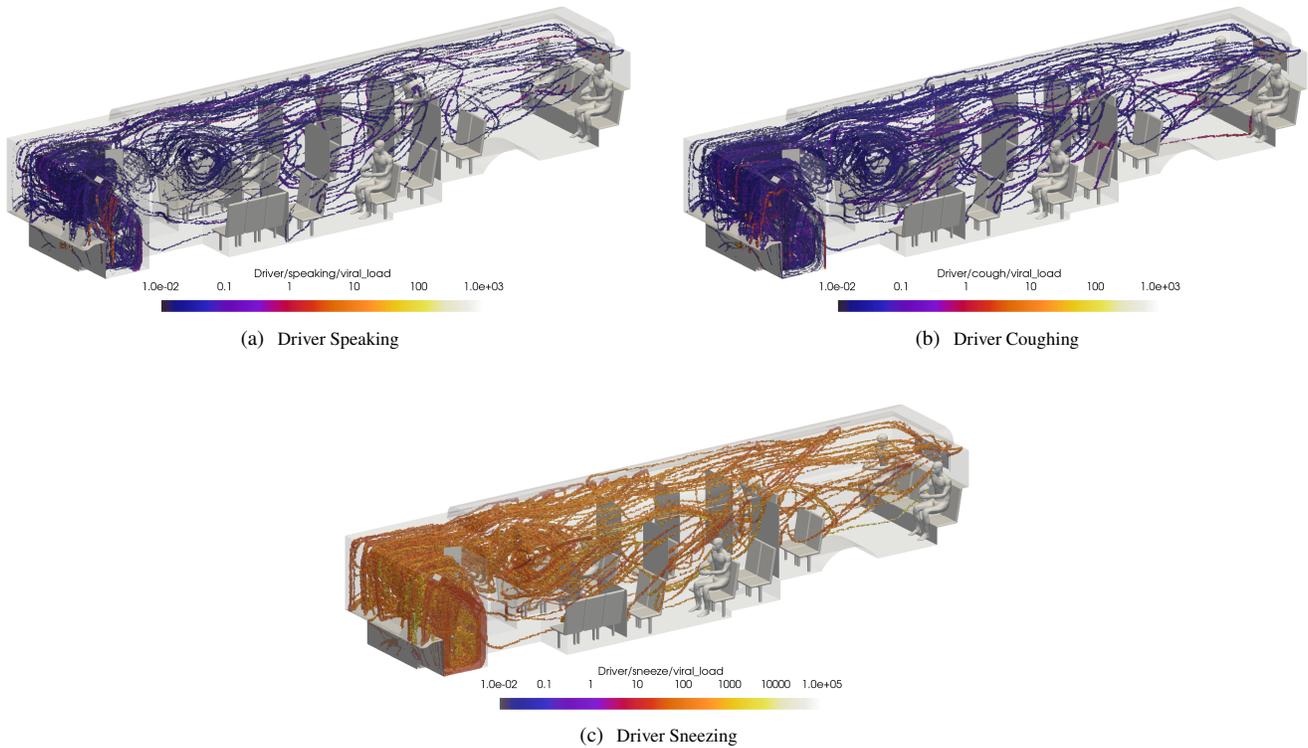

**Figure 16.** Volumetric rendering of regions of non-zero VMET for the fiducial case, comparing three injection events of the bus driver a) speaking, b) coughing, and c) sneezing.

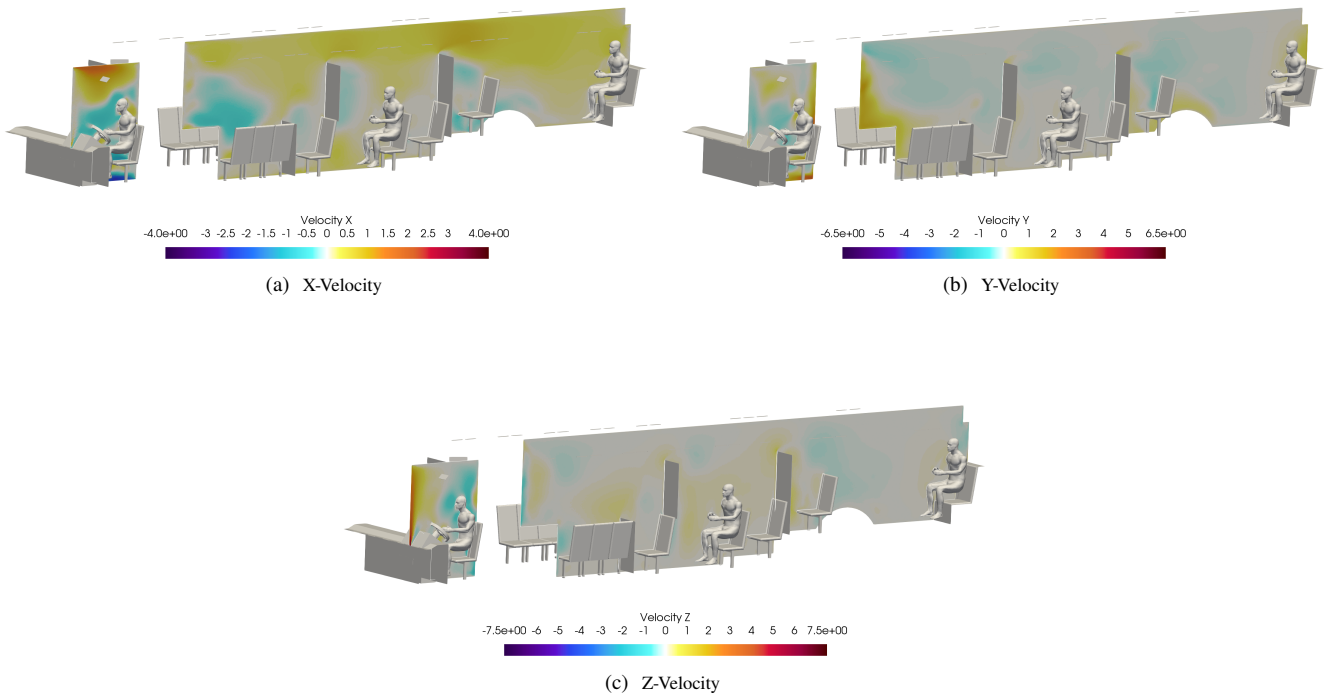

**Figure 17.** Slices of velocity taken at the bus driver location. The overall flow is toward the rear of the bus, as indicated by the positive x-direction velocity (a). The injection of air from the defroster is also clearly visible in the z-direction velocity (c).

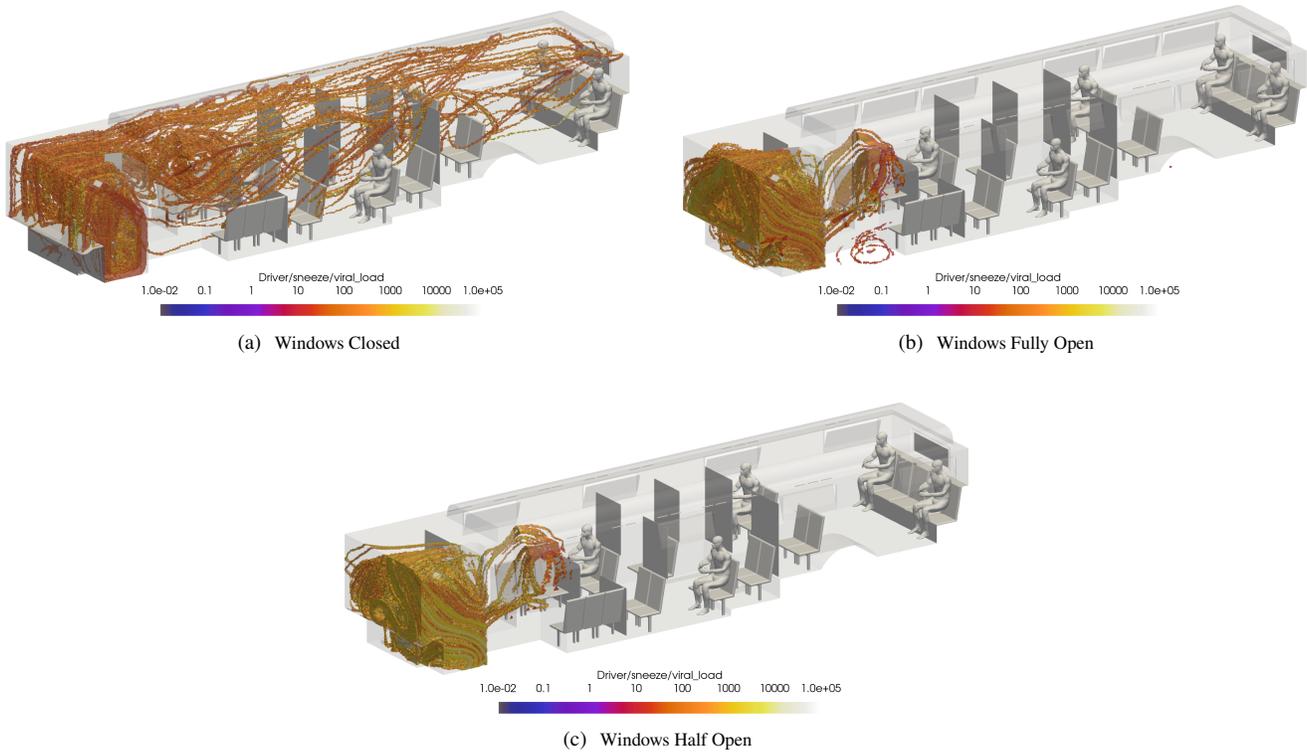

**Figure 18.** Volumetric rendering of regions of non-zero VMET comparing the dispersion of particles when the bus driver sneezes for the a) windows closed case, b) windows fully open case, and c) windows half open case.

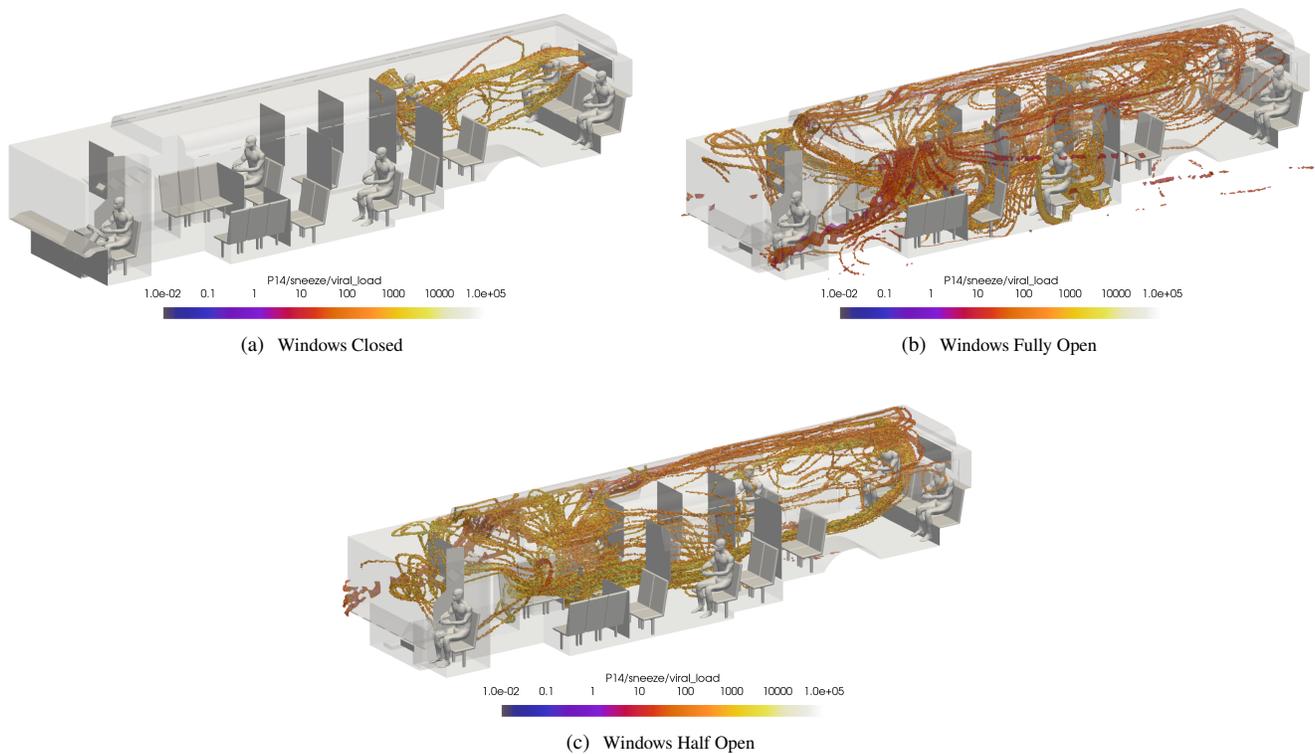

**Figure 19.** Volumetric rendering of regions of non-zero VMET comparing the dispersion of particles when passenger 14 sneezes for the a) windows closed case, b) windows fully open case, and c) windows half open case.

With the majority of cases yet to be investigated, it is difficult at this time to recommend an ideal configuration of HVAC and windows open/closed for the prevention of viral transmission. However, from these initial cases it seems that the spread of viral particles from infected passengers may be reduced with the windows closed and the cabin HVAC set to high. This is a result of the positive velocity in the x-direction directing the particles toward the main HVAC return rather than circulating the particles throughout the bus due to the increased turbulence in the flow field from open windows. In contrast, it seems that both windows open cases reduce the spread of viral particles from the driver. It is possible that keeping the majority of the windows closed, but leaving the window just rear of the driver open could lead to a best case scenario.

However, it is also possible that having the driver HVAC off contributes to the lack of circulation. To isolate this phenomena, one additional case is considered. The alternate configuration B for the windows half open case has the same HVAC settings as both previous windows open cases, but has the window behind the driver closed. Fig. 21 compares the volumetric VMET when the bus driver sneezes for the windows half open and alternate open B cases. When the window just rear of the driver is closed, particles are observed to be dispersed throughout the cabin of the bus. This demonstrates that having the window just rear of the driver open could significantly limit the spread of the driver's aerosols in most cases.

## 5 Conclusions

In this research effort our team has developed a high-fidelity database of scenarios that relate to operation in a Charlotte Area Transit System (CATS) Bus. High-fidelity Navier-Stokes solutions have been developed for a range of configurations, HVAC settings, and operational settings. An open-source Lagrangian particle tracking method was modified to include evaporative effects and used to simulate the trajectories of respiratory particles injected into the flow via speaking, coughing, and sneezing. The trajectories were then mapped to quantify the average risk of exposure to the SARS-CoV-2 virus across all considered flow scenarios. This viral exposure map provides insight into the common areas of virus collection across all configurations and settings which may be leveraged to better understand viral transmission and thus establish procedural changes that can mitigate the spread of COVID-19, and other aerosolized virus particles.

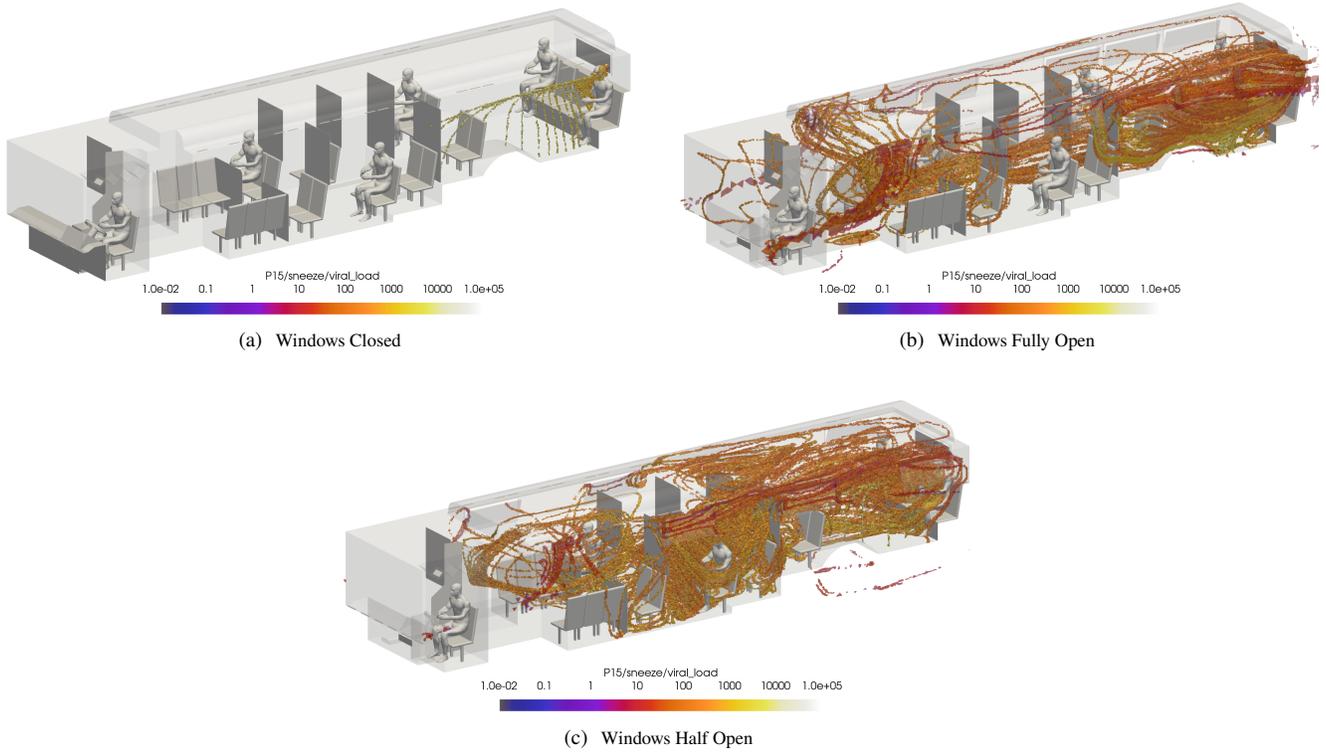

**Figure 20.** Volumetric rendering of regions of non-zero VMET comparing the dispersion of particles when passenger 15 sneezes for the a) windows closed case, b) windows fully open case, and c) windows half open case.

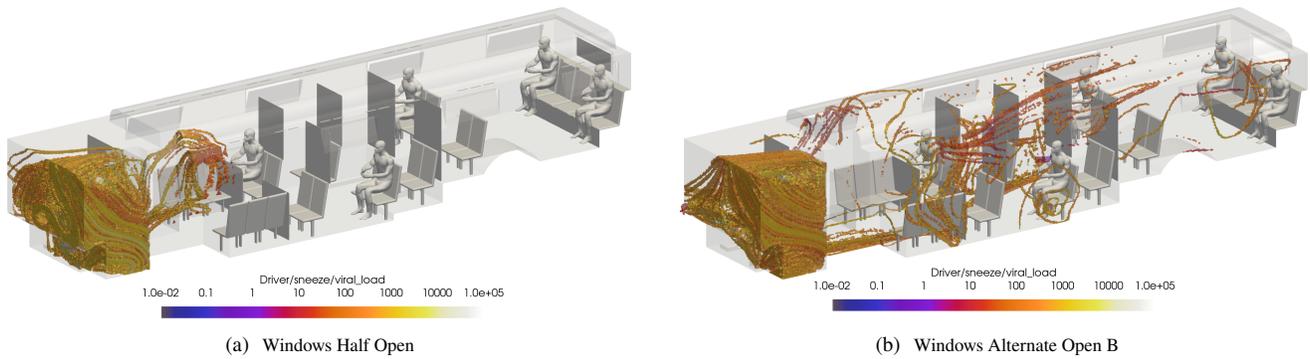

**Figure 21.** Volumetric rendering of regions of non-zero VMET comparing the dispersion of particles when the bus driver sneezes for the a) windows half open and b) windows alternate open B.

**Acknowledgements**

This project is supported by the Coronavirus Aid, Relief, and Economic and Security Act (CARES Act) as part of an award from the North Carolina Pandemic Recovery Office. The contents are those of the author(s) and do not necessarily represent the official views of, nor an endorsement, by the State of North Carolina or the U.S. Government. All CFD Computations were performed on the high-performance computing cluster at Corvid Technologies.